# Would Monetary Incentives to COVID-19 vaccination reduce motivation?

Eiji Yamamura[1*], Yoshiro Tsutsui[2], Fumio Ohtake[3],

[1] Department of Economics, Seinan Gakuin University, Fukuoka, Japan

[2] Kyoto-Bunkyo University, Japan

[3] Osaka University, Japan

*Corresponding author

Email: yamaei@seinan-gu.ac.jp

# Abstract

Some people did not get the COVID-19 vaccine even though it was offered at no cost. A monetary incentive might lead people to vaccinate, although existing studies have provided different findings about this effect. We investigate how monetary incentives differ according to individual characteristics. Using panel data with online experiments, we found that (1) subsidies reduced vaccine intention but increased it after controlling heterogeneity; (2) the stronger the social image against the vaccination, the lower the monetary incentive; and (3) persistently unvaccinated people would intend to be vaccinated only if a large subsidy was provided.

# 1. Introduction

Even after COVID-19 vaccines were made available, many people chose not to get vaccinated. A monetary incentive was anticipated to improve health behaviors even before the appearance of COVID-19 (Giles et al. 2014); more specifically, it might be effective to increase the rate of COVID-19 vaccination (Jecker 2021). If people were compensated, most of them would intend to get the shot of the COVID-19 vaccine, whereas 14% of people would not get vaccinated (Carpio et al. 2021). Then, a monetary incentive such as a subsidy is recommended to increase the rate of vaccination (Qin, Wang, and Ni 2021; Wang et al. 2021). As policies to provide a monetary incentive have been implemented for emergency situations, the impact of a monetary incentive has been investigated in many studies such as survey experiments (Campos-Mercade et al. 2021; Klüver, Humphreys, and Giesecke 2021), field experiment (Jacobson et al. 2022), policy using lottery programs (e.g., Acharya and Dhakal 2021; Barber and West 2022; Brehm, Brehm, and Saavedra 2022; Dave et al. 2021; Law et al. 2022; Sehgal 2021; Walkey, Law, and Bosch 2021), and a comparison between lottery and guaranteed rewards (Thirumurthy et al. 2022).

These studies have offered conflicting accounts of the effectiveness of monetary incentives. The anticipated effect has not been observed in many studies (Dave et al. 2021; Jacobson et al. 2022; Law et al. 2022), whereas some have found it effective to increase the vaccine uptake (Acharya and Dhakal 2021; Campos-Mercade et al. 2021). Other studies have reported mixed results (Thirumurthy et al. 2022; Walkey et al. 2021). In contrast, some studies have found that financial incentives may backfire (Hsieh 2021; Lowenstein and Cryder 2020).

The reluctance of people to obtain available vaccines has been known as "vaccine hesitancy" (MacDonald 2015), and it has been addressed by researchers to analyze the reasons for which the vaccination rate did not rise (Brita Roy, Vineet Kumar, and Arjun Venkatesh 2020; Doherty et al. 2021; Kelekar et al. 2021; Lucia, Kelekar, and Afonso 2021). The reasons to be hesitant toward vaccination are various. Considering the heterogeneity of individuals is critical to investigate the effectiveness of monetary incentives, and policymakers should provide a monetary incentive to target subjects who would be more likely to get the shot if given incentives. Some studies have paid some attention to heterogeneity and found the incentive's effect to differ across political views (Jacobson et al. 2022; Thirumurthy et al. 2022). However, existing studies have not scrutinized heterogeneity enough to consider the monetary incentive.

Psychological or ethical factors should be taken into account to explore why individuals are reluctant to get vaccinated (Murphy et al. 2021); for instance, people would become less willing to take the vaccine if they considered the monetary incentive inappropriate for human life and a social problem. Introducing a monetary incentive may result in an unexpected outcome when social norm is considered (Gneezy and Rustichini 2000). According to the self-image hypothesis derived from intrinsic motivation, prosocial motivation is a key factor in determining an individual's behavior (Benabou and Tirole 2003; Bénabou and Tirole 2006). Thus, intrinsic prosocial motivation reasonably leads people to be reluctant to vaccinate if the monetary incentive goes against their motivation.

Economists have recently drawn attention to "social image," which also provides clues to analyze the reason why the vaccination rate did not rise. In schools, students may want to show that they are "cool" to their classmates, and they would avoid making an educational investment if the classmates consider it "cool." This may explain why the supporters of Donald Trump tend to be unvaccinated (Jacobson et al. 2022)[1]: they are incentivized to maintain their social image (Bursztyn and Jensen 2017). People's behavior is to follow the norm shared in the group to which they belong even under an emergency situation such as COVID-19 (Takahashi and Tanaka 2021). Inevitably, a monetary incentive seems to have different effects according to people's characteristics, norm, motivation, and social image. Considering heterogeneity, its positive and negative effects may have been neutralized in many studies and thus could not explain the effect of a monetary incentive on the uptake of COVID-19 vaccine. However, no study has been conducted that addresses this problem.

---

[1] Social image influenced voting behavior under elections considered a natural experiment (Dellavigna et al. 2017; Funk 2010; Gerber, Green, and Larimer 2008; Jacobson et al. 2022).

    This study aimed to examine the effect of a monetary incentive on vaccine intention using monthly individual-level panel data through Internet surveys in Japan using a sample of about 93,000 people. We found that (1) hypothetical subsidies for vaccination reduced vaccine intention; (2) the subsidy increased intention once heterogeneity was controlled; (3) the subsidy effect was lower for those with prosocial motivation than others; (4) the social image of being against vaccination reduced intention; and (5) a subsidy of 5,000 yen was effective for unvaccinated people (20% of the population), but 1,000 yen were not.

## 2. Hypotheses

    Some behaviors could not be appropriately incentivized through a market mechanism partly because behaviors are not ethically admitted. For instance, organ trafficking should

not be promoted in the market because putting price on organs is against ethics and social norm. This also holds for vaccination, which is related to human life and for which social norm should be followed.

Under the COVID-19 pandemic, people were in a predicament and exposed to life-threatening danger. In this scenario, COVID-19 vaccines were developed to save lives. People were urged to take the vaccine to protect not only their own lives but also those of others by mitigating the spread. Because of life-related issues, people might argue that monetary incentives should not be provided to encourage vaccination. If people receive such incentives, they might be punished by members of society because they are regarded as norm breakers, especially under emergency situations such as COVID-19 (Takahashi and Tanaka 2021).

Here we propose *Hypothesis 1*:

*Hypothesis 1: Monetary incentives reduce vaccination intention.*

The norm might differ between groups. Therefore, people's reaction to provided monetary incentives depends on the norm shared by group to which they belong. Here we raise *Hypothesis 2*:

*Hypothesis 2: Monetary incentives have different influences on vaccine intention among social groups.*

Individual behavior is influenced by the norm shared by the group to which one belongs through peer pressure. Thus, one is motivated to keep one's social image to align with this norm (Bursztyn and Jensen 2017). Let us assume two types of groups: in one group, members share the norm to be vaccinated, encouraging vaccine uptake; in another group, members share the norm to be against vaccination, discouraging vaccine uptake. Further, social image is strengthened when the group to which one belongs becomes relatively smaller in society (Abramitzky, Einav, and Rigbi 2010). As the negative effect of social image increases, the positive effect of a monetary incentive would be neutralized. This leads us to propose *Hypothesis 3*:

*Hypothesis 3: As the rate of vaccination increases, a monetary incentive for unvaccinated people becomes less effective.*

## 3. Data and Methods

### 3.1. Data collection

At the beginning of March, directly after COVID-19 reached Japan, we planned to collect and construct the panel data. At that time, a COVID-19 vaccine had not yet been developed; this is why the sampling method could not be specifically and purposefully designed to investigate issues of COVID-19 vaccination. Nonetheless, we started by investigating identical individuals almost every month through repeated surveys. Subsequently, after these individuals had been offered the COVID-19 vaccine, questions about vaccination were included in the questionnaire.

The internet surveys were outsourced to the research company INTAGE, Inc., which has extensive experience in conducting online academic surveys and has a high reputation. Many people were registered with INTAGE prior to the survey, and among them, respondents were recruited to participate. The first survey was conducted on March 13–16, 2020, and the rate of completion reached 54.7%.

In our sampling method design, we collected representatives of the Japanese adult population by considering various characteristics such as age, educational background, gender, and residential prefectures. However, it should be noted that persons below 17 and over 78 years of age were excluded from the sample because they were too young and too old to properly answers the questions. Then, participants were randomly selected to fill the pre-specified quotas, and they were offered monetary incentives if they completed the survey.

**< Fig 1. Changes in percentage of vaccinated people.** This is the Figure legend.>

Internet surveys were conducted 26 times ("waves") about every month. Fig.1 demonstrates the change of percentage among those who were vaccinated by comparing official data and our data. In April 2021, directly after implementing the vaccine, most people could not get the shot due to scarcity of supply; nevertheless, in the study sample, the percentage drastically rose and reached over 80% in November 2021. However, as its marginal increase reduced, the rate of vaccination hardly increased from October to November 2021. This reflects that part of the people were reluctant about getting vaccinated. A similar increasing trend was also observed for the official data, although the vaccination rate was lower than that of the studied data.

Our simple experiments were included for eight months, from April 11, 2021 to November 18, 2021. In the November 18 survey, around 80% people were vaccinated, and the marginal increase of vaccinated respondents became very small. Table 1 shows the observations for each survey.

**< Table 1. Time point of surveys and observations.** This is the Table legend.>

We used the balanced panel to pursue identical persons by controlling for different stages because the rate of vaccinated people had rapidly risen, as demonstrated in Fig 1. During the study period, most persons who had been unvaccinated at the time of the first survey later got vaccinated; however, about 20% of participants remained unvaccinated.

## 3.2. Measurements

< **Table 2. Definitions of key variables and its mean, standard errors, and maximum and minimum values.** This is the Table legend.>

As indicated in Table 2, respondents were asked to indicate their vaccination intention on a scale from 1 (*very low*) to 5 (*very high*). As a simple experiment, all respondents were asked the same question three times in different situations. First, they only answered the question. Next, a monetary incentive was provided, and they were requested to answer the question under the following condition:

*"One can receive 1,000 yens subsidy if one gets a shot of the COVID-19 vaccine."*[2]

Third, they answered the question when the subsidy amount increased:

*"One can receive 5,000 yens subsidy if one gets a shot of the COVID-19 vaccine."*

The monetary amount became five times larger than in the former situation—an effect that has been also considered in a previous work (Jacobson et al. 2022). In this way, we examined the effect of a monetary incentive and its difference in amount for the same individuals. The same participants gave three replies under three different hypothetical conditions. For constructing the dataset, to consider the effect of the different conditions, the same respondents appeared three times in the survey when they answered the question about vaccination intention, and 3,887 identical respondents participated in the surveys from April 11 to November 18. In each survey, first, we gathered the three replies about vaccination intention for the same individuals under the three different hypothetical conditions, although other variables had the same values; hence, 3,887 observations were multiplied by 3, reaching 11,661. Subsequently, we aggregated data of surveys from the April 11 to the November 18 to construct the balanced panel; 11,661 were multiplied by 8, reaching a total of 93,288 observations. Table 2 indicates that the mean values of the dummies for "*Subsidy 1000*" and "*Subsidy 5000*" are 0.33. This reflects the structure of the dataset used in this study because the same person appeared three times by indicating vaccination intention if no subsidy, 1,000 yen subsidy, or 5,000 subsidy was offered.

---

[2] US $ 1= JP ¥ 110 on average in this period.

Individuals might have different motivation for vaccination. Considering people's heterogeneity, we also investigated how the effect of a monetary incentive varies according to different motivations and views about the vaccine's effects. For this purpose, we asked motivation for COVID-19 vaccination and views about it only in the April 11 survey, when we first included questions about vaccination. Hence, variables about it were fixed throughout the survey. As indicated in Table 1, respondents were asked to respond to five questions on a scale from 1 (*strongly disagree*) to 5 (*strongly agree*).

Regarding motivation, they were asked two questions as follows:

*"In deciding whether to get the COVID-19 vaccine, it is important to save your life."*

*"In deciding whether to get the COVID-19 vaccine, it is important that it prevents the spread of COVID-19."*

From the answers to the above questions, we obtained two variables of vaccination motivation, namely "*Self_motivation*" and "*Prosocial_motivation.*"

With regard to views about COVID-19, the respondents were asked two questions as follows:

*"The COVID-19 vaccination has a negative impact on your health."*

*"The COVID-19 vaccination is ineffective."*

From the answers to the above questions, we obtained three variables of views about vaccination motivation, namely "*Vaccine_harm*" and "*Vaccine_Ineffective*," which are considered to indicate the degree of doubt about the vaccination.

## 3.3. Descriptive statistics

As shown in Table 2, the mean value of *Vaccine Intention* is 3.88; thus, people were likely to intend to get the COVID-19 vaccine. The mean value of *No_vaccine* is 0.69, indicating that nearly 70% of people were not vaccinated during the study period on average. However, during that same period, some initially unvaccinated people received the vaccine and became vaccinated persons; accordingly, *No_vaccine* is not constant even for the same respondents. *Rate vaccine* is the rate of vaccinated people in the residential prefecture, and its mean value is 0.24, which is almost consistent with *No_vaccine*. Concerning motivation, the mean values of *Self_motivation* and *Prosocial_motivation* are over 4, indicating that most people tended to have motivation to protect themselves and to prevent the spread of COVID-19. In contrast, the mean values of *Vaccine_harm* and *Vaccine_Ineffective* are about 2.80, and their standard errors were relatively small. As for effectiveness and harm about COVID-19, most peoples' views were neutral.

<Fig 2. Comparison between vaccinated and unvaccinated respondents. This is the figure legend.>

Fig. 2 compares the motivation for "*Vaccine_intention*" and views about vaccines between vaccinated and unvaccinated respondents. People's reluctance to receive available vaccines is called "vaccine hesitancy" (MacDonald 2015). The reason why people hesitated to be vaccinated is critical, and psychological factors should be considered to investigate it (Murphy et al. 2021). As indicated by previous research (Arshad et al. 2021; Han et al. 2021; Kabir et al. 2021), subjective view about the side effects and effectiveness of the vaccine is associated with hesitancy.

Vaccinated people have higher self and prosocial motivation than unvaccinated ones. During the study period, in the early stage of implementation of the COVID-19 vaccine, the government restricted the initial group receiving the shot to health workers and older people aged 65 and over. Accordingly, 75% of older persons were vaccinated in July 2021 (Japan Times 2021). However, as time passed, anybody could get the shot, and motivation to be vaccinated became critical. Meanwhile, unvaccinated people were more likely to consider the vaccine to be harmful and ineffective. As a whole, these findings are in line with those of previous studies: the motivation for accepting COVID-19 was personal protection against COVID-19, whereas hesitancy was commonly explained by concerns about the vaccine's side effects (Almaghaslah et al. 2021; Machingaidze and Wiysonge 2021; Solís Arce et al. 2021).

## 3.4. Methods

We used a fixed-effects (FE) model to control the time-invariant individual characteristics. The estimated baseline function takes the following form:

$$\text{Vaccine intention}_{it} = \alpha_1 \text{ Subsidy 1000}_{it} + \alpha_2 \text{ Subsidy 5000}_{it} + \alpha_3 \text{ No\_vaccine}_{it} + \alpha_4 \text{ Rate Vaccine} + A'X + k_t + m_i + u_{itg},$$

In this formula, $\text{Vaccine intention}_{it}$ represents the dependent variable for individual $i$ and wave $t$. The regression parameters are denoted as α, and $m_i$ represents time-invariant fixed effect. The FE model controls various individual characteristics that are fixed even if time has passed; therefore, it controls for gender, growth environment, parents' characteristics, educational background, and various factors. $k_t$ represents the effects of different time points, which are various shocks that occurred simultaneously throughout Japan at each time point. Seven survey dummies were included to control $k_t$, and the error term is denoted as $u$. We conducted estimation using the full sample consisting of vaccinated and unvaccinated people and a subsample of unvaccinated people.

$\text{Subsidy 1000}_{it}$ and $\text{Subsidy 5000}_{it}$ are dummies for the condition where a subsidy is provided to get the shot. The key independent variables are the cross terms between subjective variables and subsidy dummies. We included $\text{No\_vaccine}_{it}$, which is a dummy for those who have not been vaccinated. Unvaccinated people were considered to be more hesitant to take the vaccine than vaccinated ones; thus, the expected sign of $\text{No\_vaccine}_{it}$ is negative. We investigated how the monetary incentive changed according to social image. As discussed in section 3, the social image of being against vaccination is expected to be strengthened by increases in the rate of vaccinated people. To test *Hypothesis 3* using a subsample of unvaccinated people, *Rate vaccine* was included.

In the function, we included cross terms between subsidy dummies and various variables to reflect people's heterogeneity. Those who had been vaccinated without any subsidy were likely to feel injustice if a subsidy was provided for vaccination, and the provision of a subsidy would reduce their incentive to get the shot. Thus, compared with vaccinated people, a monetary incentive gave unvaccinated people greater motivation to take the vaccine, and *No_vaccine$_{it}$* ×*Subsidy 1000* and *No_vaccine$_{it}$* ×*Subsidy 1000* were expected to be positive. Further, the difference between their coefficients indicates how effect of 1,000 yen and that of 5,000 yen are different. This holds for other interaction terms with subsidy dummies.

To consider how the monetary incentive varied according to the heterogeneity of motivation, we included *Self_motivation* in the interaction with the subsidy dummy. Its coefficients indicated whether the effect of subsidy on the intention to be vaccinated increased if people tended to be motivated to protect their life. Apart from it, introducing a monetary incentive would change the aim of vaccination (Gneezy and Rustichini 2000).

Subsidy is regarded as a part of income; thus, providing a subsidy increase individual utility while hurting its self-image of serving society by mitigating COVID-19. Hence, a monetary incentive reduced prosocial motivation. To test this conjecture, *Prosocial motivation* was included in the interaction with the subsidy dummies. These coefficients of the interaction terms between *Prosocial* and the subsidy dummies were expected to have the negative sign because an increase in self-interest from the subsidy would substitute prosocial motivation.

Concerns about side effects and the ineffectiveness of COVID-19 vaccination commonly explained people's hesitancy toward it (Almaghaslah et al. 2021; Machingaidze and Wiysonge 2021; Solís Arce et al. 2021). Therefore, the interaction terms between *Vaccine harms (Vaccine ineffective)* and subsidy dummies were included. As the concerns might be compensated by the monetary incentive, we predicted that their coefficient would take the positive sign.

As touched upon earlier, using a subsample of those who were not vaccinated, *Rate vaccine* was included in the interaction with the subsidy dummies to examine the effects of social image on vaccine intention (Bursztyn and Jensen 2017). Here we assume that two groups exist in a society: in one group, members are encouraged to take the COVID-19 vaccine for prevention, whereas in the another, members are led to refuse it.

Unvaccinated persons in the former group have a greater incentive to be vaccinated to maintain their social image if the rate of vaccinated people rises. On the contrary, in the latter group, pressure from other members incentivizes its members to continue to be unvaccinated. The latter is the group of those who are vaccine hesitant (Jacobson et al. 2022; Thirumurthy et al. 2022); if this holds true, the rise in vaccination rate strengthened their pressure to be against the vaccine because the norm of minority was strongly enforced so they would maintain their identity (Abramitzky et al. 2010). We observe this mechanism among supporters of Donald Trump in the United States, who are more likely to be vaccine hesitant (Jacobson et al. 2022). This might be one reason why the rate of vaccinated people struggles to increase. Based on "social image" theory, the sign of *Rate vaccine* × *Subsidy 1000 (or Subsidy 5000)* was predicted to be negative.

## 4. Results

### 4.1. Full sample estimations

< **Table 3**. **Dependent variable is Vaccine Intention. FE model using balanced panel sample.** This is the Table legend.>

Tables 3 reports the estimation results of the baseline model based on the whole sample. We begin by examining column (1). The coefficients of *Subsidy 1000* and *Subsidy 5000* show the negative sign and are statistically significant at the 1% level. Contrary to the assumption of rational individuals in standard economics, people are less likely to intend to be vaccinated if monetary incentives are provided than if such incentives are not given. One possible explanation is that monetary incentives reduce other incentives, such as maintaining one's self-image and social image. This result strongly supports *Hypothesis 1*. The scale of their coefficients reveals that the negative effects of subsidy become smaller as the subsidy amount increases. The positive effect of the increase in amount through the income effect reduces the negative effect. In columns (2) and (3), the sign of *Subsidy 1000* and *Subsidy 5000* changed to positive after heterogeneities were controlled by including interaction terms. This implies that monetary incentives increased vaccine intention if an individual's subjective motivation and views about the vaccine were controlled. In other words, heterogeneity plays a critical role in promoting monetary incentives.

The significant negative sign of *No_vaccine* reflects that unvaccinated people are less likely to have vaccine intention, which is consistent with our intuition. The sign of *Rate vaccine* is significantly negative in column (1) but positive in column (2). Further, *Rate vaccine* does not show statistical significance in column (3). Thus, the results about the rate of vaccinated people in a residential prefecture are not stable.

Let us now consider the results of the interaction terms. First, we observe the positive sign for *No_vaccine×Subsidy 1000* and *No_vaccine×Subsidy 5000,* with a statistical significance at the 1% level. This implies that monetary motivation is greater for unvaccinated people than vaccinated ones. *Rate vaccine× Subsidy 1000* and *Rate vaccine× Subsidy 5000* show significant negative signs in three columns even after controlling other individuals' heterogeneities, including the experience of COVID-19 vaccination. The increase in vaccination rate in residential areas reduced the monetary incentive, and one possible interpretation is as follows. Most people were vaccinated without receiving a subsidy; therefore, they consider the provision of a subsidy to be unfair because they could not receive it when they got their own shot. For a closer examination of the conjecture about social image proposed in section 3, we examine the results of Table 4 on a subsample of unvaccinated people.

The interaction terms of *Prosocial motivation* had statistical significance and distinctly changed the monetary incentive effect. Significant negative signs for *Prosocial motivation×Subsidy 1000* and *Prosocial motivation×Subsidy 5000* are observed in all results, which reflects that prosocial motivation would decrease if monetary incentives were provided. In our interpretation, if a monetary incentive is offered, its self-image of serving society is hurt. Considering the results shown in Table 3 leads to support *Hypothesis 2*.

The linear combination effect of subsidy variables and their interaction terms are visualized in Figs. 3. Let us now consider the two bars on the left, which show the effect based on the full sample. With the exception of Fig. 3 (c), the linear combination shows the positive effect on vaccine intention at the 5% level, although no significant difference is observed between *Subsidy 1000* and *Subsidy 5000*. The insignificance of Fig. 3 (c) reflects that a monetary incentive is ineffective for those with prosocial motivation, which is consistent with the argument of self-image (Benabou and Tirole 2003; Bénabou and Tirole 2006).

< **Figs. 3. Effect of subsidy using linear combination.** This is the Figure legend.>

## 4.2.　Subsample estimations

< **Table 4**. **Dependent variable is Vaccine Intention. FE model using sub-sample of unvaccinated people.** This is the Table legend.>

We turn to Table 4. In columns (1) and (2), the results are based on a subsample of unvaccinated people at each time point of the surveys. Meanwhile, in columns (3) and (4), the results are on the subsample of unvaccinated people in the last survey conducted in November 2021. As 7 months had passed since the vaccine became available (see Fig. 1), they can be considered the group sharing the norm against the vaccine; that is, they were unlikely to get the shot in order to maintain their social image with group members. The specification of the estimated model differed from that in Table 3 in that "*No-vaccine*" and its interaction term are not included because unvaccinated people are not included.

*Subsidy 1000* and *Subsidy 5000* show similar results to those of Table 3. However, we should pay a careful attention to column (3). *Subsidy 5000* shows the positive sign and statistical significance at the 1% level even when heterogeneities are not controlled. That is, persistently unvaccinated people would change their vaccine intention if the monetary incentive became large, regardless of heterogeneities.

Figs. 3 demonstrate that *Subsidy 5000* significantly increased vaccine intention, whereas *Subsidy 1000* did not show statistical significance. The exceptional case is in Fig. 3(a), where any monetary incentive did not increase intention if the vaccination rate was considered. That is, as shown in Table 3, social image shared by persistently unvaccinated people reduced the monetary incentive; hence, positive monetary incentives would be neutralized.

**< Figs. 4. Effect of vaccination rate using linear combination. "Rate vaccine +Subsidy×Rate vaccine"**. This is the Figure legend.>

Fig. 4 demonstrates the effect of the rise in vaccination rate on vaccine intention. In the results on the full sample, the influence is small, and statistical significance is not so high, while in the subsample of persistently unvaccinated people, the rise in vaccination rate distinctly reduced intention regardless of the subsidy amount. The combined results of Table 4 and Fig. 4 are consistent with *Hypothesis 3*. In our interpretation, social image has a large negative effect and outweighs the positive effect of monetary incentives.

## 5. Discussion

In the United States, monetary incentives increased vaccination intention but not the vaccination rate (Jacobson et al. 2022). In contrast, we found that a monetary incentive reduced vaccination intention if individuals' heterogeneity was not considered, and the effect of the monetary incentive increased once heterogeneity was controlled.

The findings of this study have policy implications as follows. Receiving a subsidy is thought to contradict its intrinsic self-image, leading to a significant reduction in positive monetary incentive; accordingly, the monetary incentive is neutralized. The key implication of this study is that vaccine intention is opposed to the key implication of traditional economics that pursuing self-interest increases social welfare. Therefore, rational behavior for self-interest does not contradict a prosocial outcome.

However, the self-image or social image to be maintained by following the norm might differ among individuals. Considering heterogeneity, if a small monetary incentive was provided, unvaccinated people would intend to be hesitant to get vaccinated compared to vaccinated people. However, monetary incentives would increase their intention if the subsidy amount was large enough. Further, the impact of monetary incentives depends on the subsidy amount. Meanwhile, the higher the rate of vaccinated people, the lower the vaccine intention becomes, and the effect of social image in the smaller group becomes larger. Inevitably, that vaccination rate could not reach 100% even if a large monetary incentive was provided; therefore, we should attempt to rise the vaccination rate but not to 100%. Considering findings together leads policymakers to consider heterogeneity to introduce a monetary incentive but not attempt to make all people vaccinated.

The limitation of this study is twofold. First, it should be noted that the effect of the monetary incentive on behavior may differ from that on intention. Therefore, in response to the monetary incentive, how people might behave in the real world is not explored. Further, whether the findings of this study are specific to the setting of Japan is unknown.

Thus, these issues should be scrutinized.

Second, social image plays a substantial role in people's behaviors if these behaviors are known to other members of the group; that is, social image has no influence if other cannot observe their behaviors because people would not need to maintain their social image with other members. In this case, self-image is more important than social image (Akerlof and Kranton 2000; Benabou and Tirole 2003). We already controlled prosocial motivation based on self-image (Bénabou and Tirole 2006). In the setting of this study, participants' replies to the questions were not viewed by other members, and hence, our interpretation is that the respondents sought to maintain self-image rather than social image.

## 6. Conclusion

The rate of vaccinated people remained at a certain level even though the vaccines were offered at no cost in various countries. We investigated the how provision of a monetary incentive changed vaccine intention in Japan. We found that (1) providing a monetary incentive reduced vaccine intention, despite increasing it after controlling heterogeneity of self-image, social image, and concerns about the vaccine; (2) the positive effect of subsidy is neutralized by its negative effect derived from hurting the self-image of prosocial people; (3) the rate of vaccinated people increased in the residential prefecture, while the monetary incentive decreased, which is considered an effect of social image; (4) 20% of people continued to be unvaccinated even after 7 months since the vaccine was made available. A subsidy of 5,000 yen increased people's motivation, but a subsidy of 1,000 yen had no effect.

Policymakers should establish that a monetary incentive be provided only to those who continue to be unvaccinated and are hesitant to take the COVID-19 vaccine because they intend to maintain their social image against it with members of the group to which they belong. However, a monetary incentive would motivate them to get the shot if the subsidy amount was large. Nonetheless, if vaccination rate increased to a high level, the monetary incentive would become ineffective; thus, policymakers should not attempt to reach a 100% rate of vaccinated individuals, although the optimal rate is not indicated. The findings showed that the zero COVID-19 policy cannot be realized. Further, it is unknown whether a monetary incentive would actually lead unvaccinated people to get the shot even if vaccine intention increased owing to it. This should be scrutinized in future studies.

# Acknowledgments

We would like to thank Editage (http://www.editage.com) for editing and reviewing this manuscript for the English language. This study was supported by Fostering Joint International Research B (Grant No.18KK0048) and Grant-in-Aid for Scientific Research S (Grant No. 20H05632) from the Japan Society for the Promotion of Science.

## Compete of interest

There is no conflict of interest in this study.

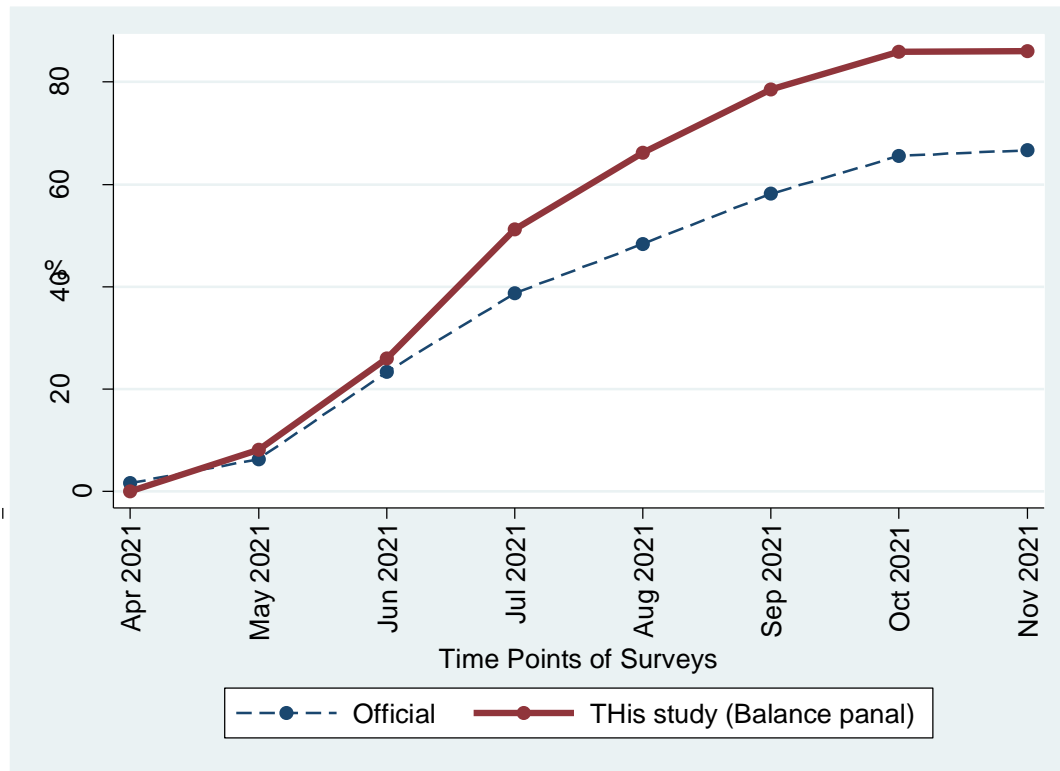

**Fig. 1 Changes in percentage of vaccinated people.**
Source. Prime Minister's Office of Japan.
https://www.kantei.go.jp/jp/headline/kansensho/vaccine.html (accessed Jan 11, 2022)

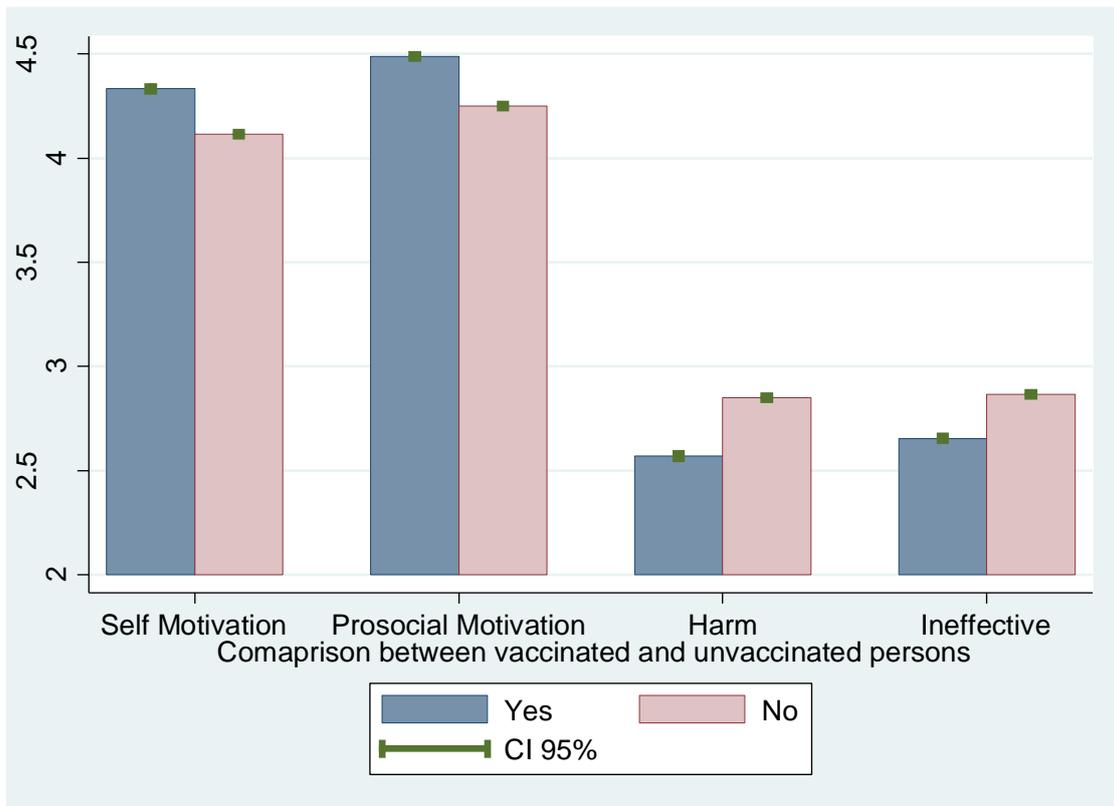

Fig. 2 Comparison of subjective values about the COVID-10 vaccine between vaccinated and unvaccinated persons.

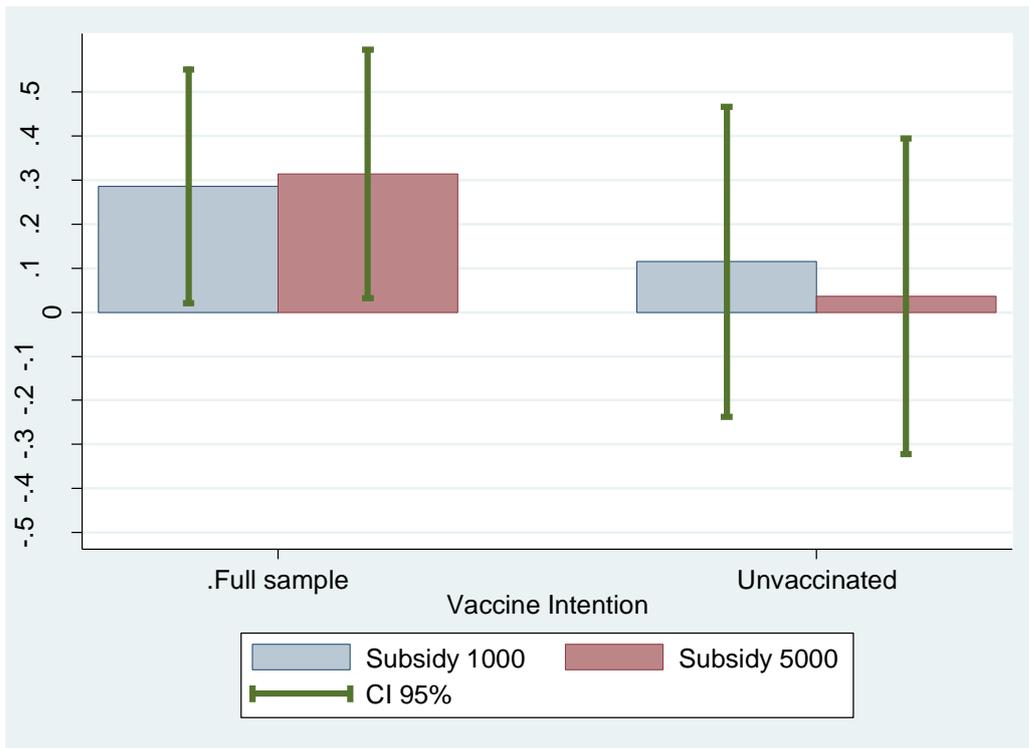

(a) Linear combination Subsidy+Subsidy×Rate vaccine.
   Column (3) of Table 3 and Column (4) of Table 4

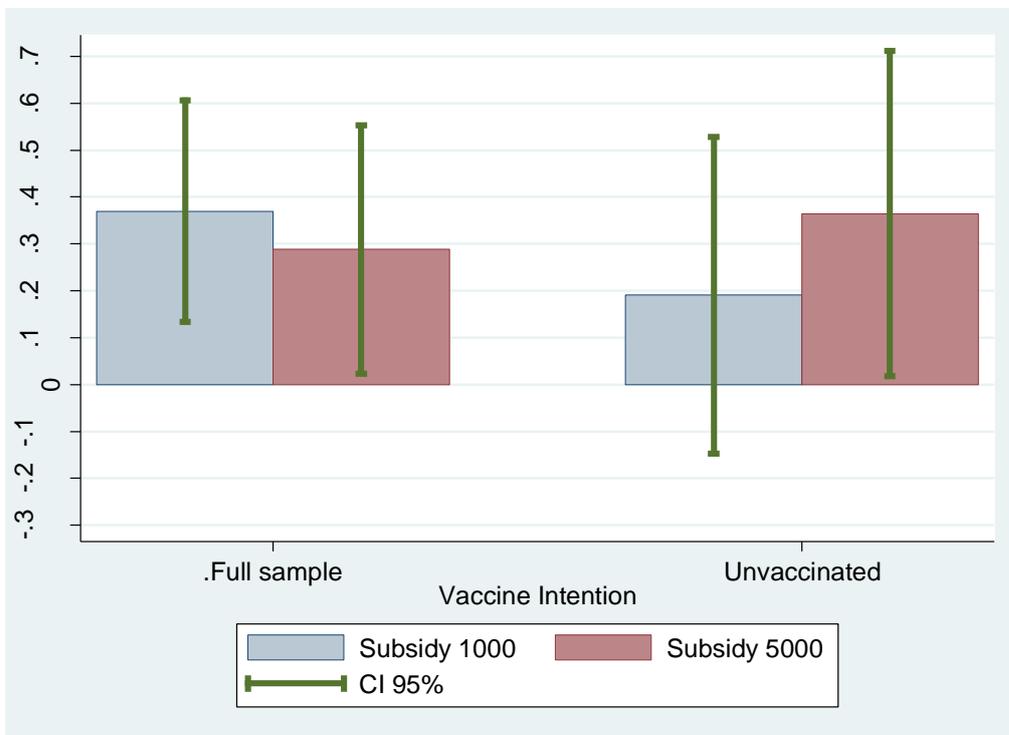

(b) Linear combination Subsidy+Subsidy×Self_motivation
   Column (3) of Table 3 and Column (4) of Table 4

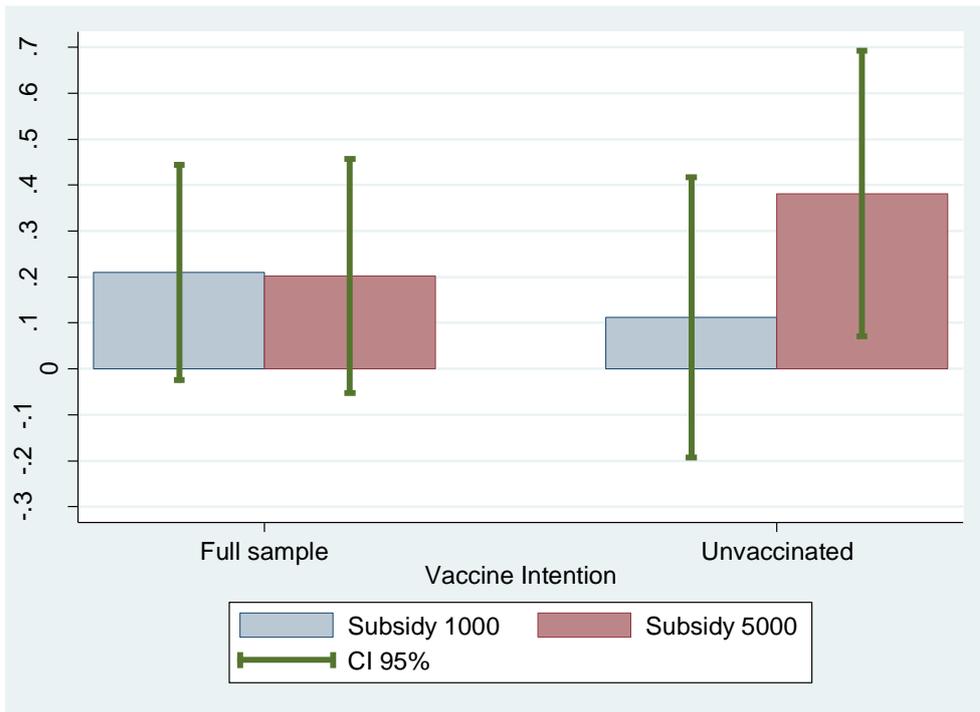

**(c) Linear combination Subsidy+Subsidy×Prosocial_motivation**
  Column (3) of Table 3 and Column (4) of Table 4

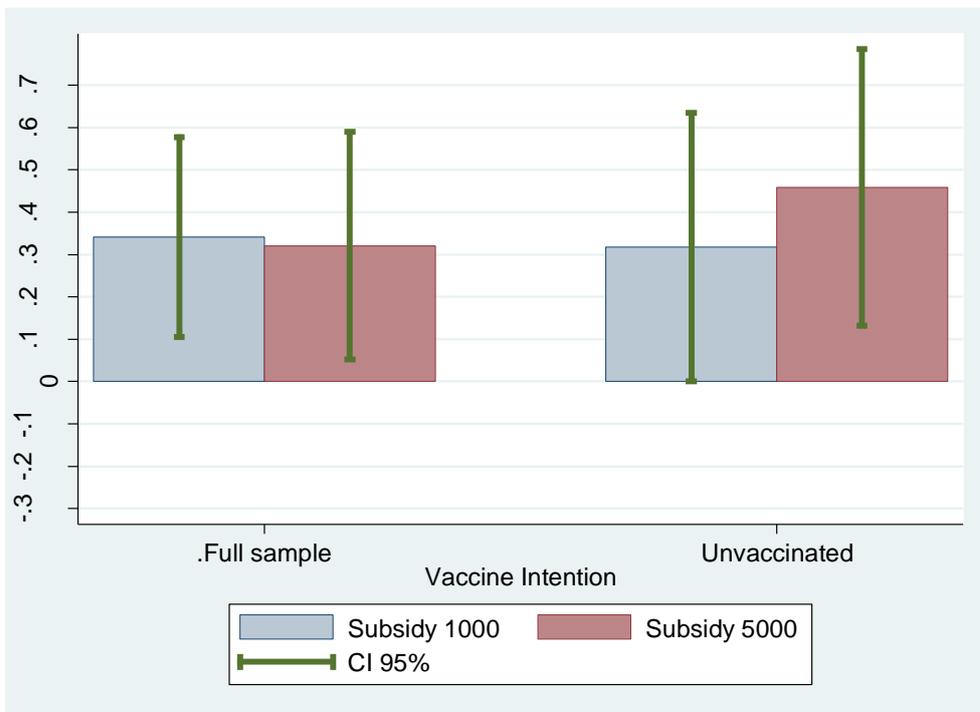

**(d) Linear combination Subsidy+Subsidy×Vaccine_harm**
  Column (3) of Table 3 and Column (4) of Table 4

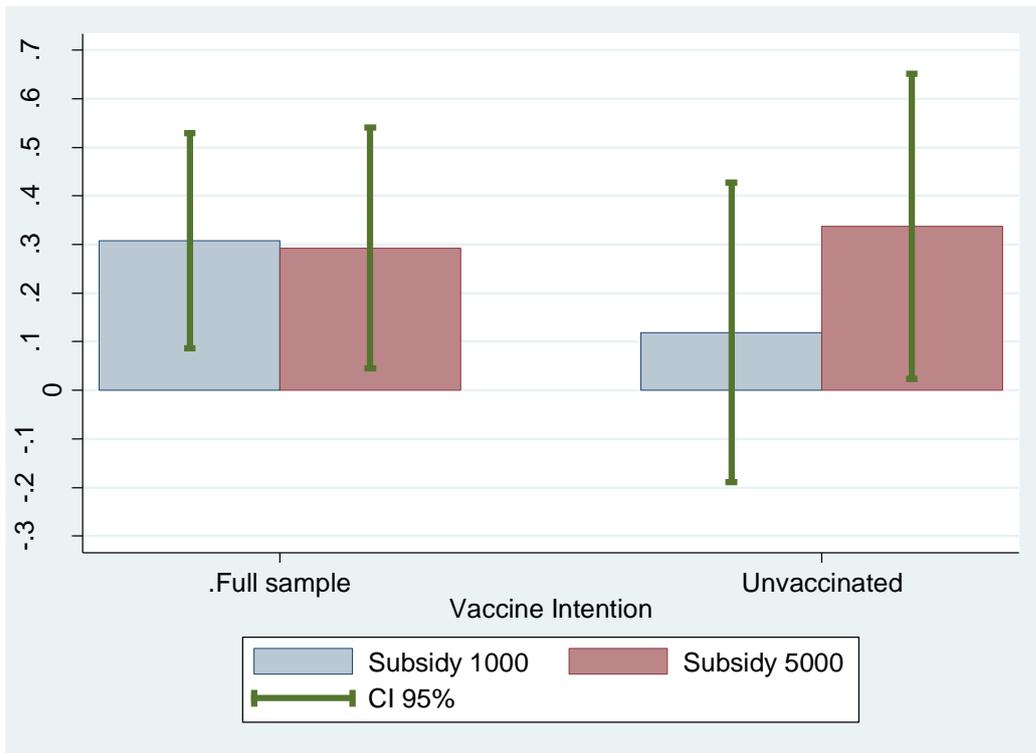

**(e) Linear combination Subsidy+Subsidy×Vaccine_Ineffective
Column (3) of Table 3 and Column (4) of Table 4**

**Fig. 3. Effect of subsidy using linear combination.**

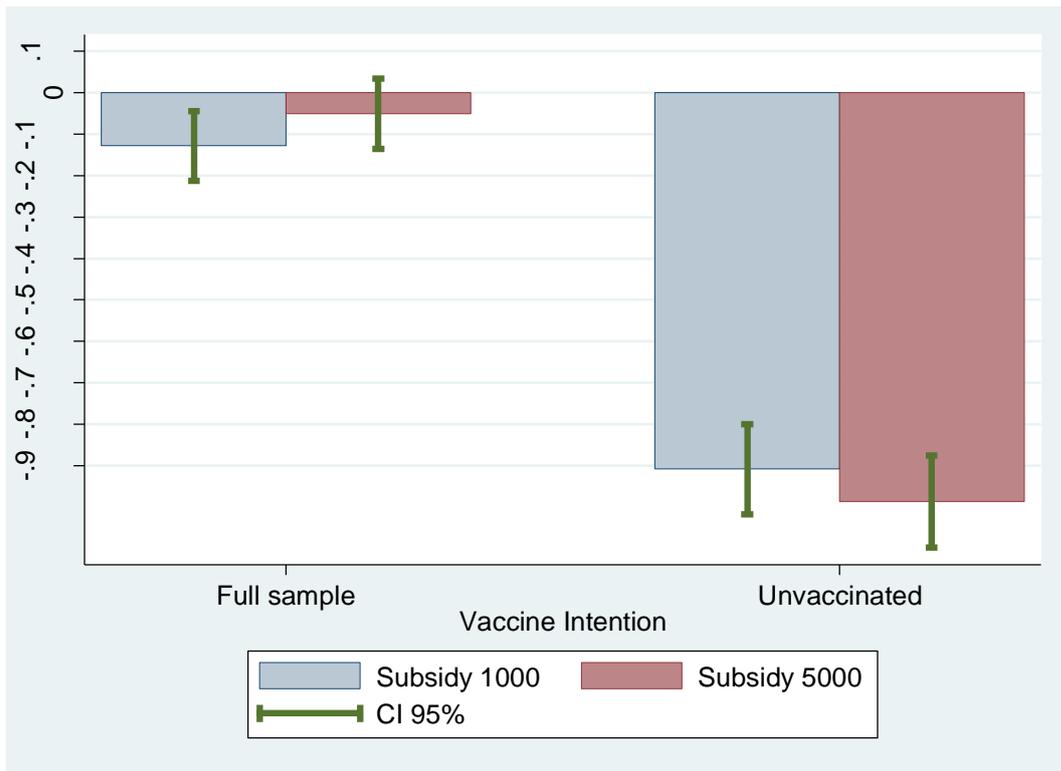

**Effect of vaccination rate using linear combination. "Rate vaccine +Subsidy×Rate vaccine."**

**Column (3) of Table 3 and Column (4) of Table 4**

**Fig. 4. Effect of social image. Marginal effect of vaccine rate in residential prefecture on the vaccine intention using linear combination.**

**Table 1. Time point of surveys and observations.**

| Number of Surveys | Dates | Obs. |
|---|---|---|
| 1 | March 13–16, 2020 | 4,359 |
| 2 | March 27–30, 2020 | 3,495 |
| 3 | Apr 10–13, 2020 | 4,013 |
| 4 | May 8–11, 2020 | 3,996 |
| 5 | June 12–15, 2020 | 3,877 |
| 6 | Oct 23–28, 2020 | 3,626 |
| 7 | Dec 4–8, 2020 | 3,491 |
| 8 | Jan 15–19, 2021 | 3,509 |
| 9 | Feb 17–22, 2021 | 3,529 |
| 10 | Mar 24–29, 2021 | 3,440 |
| 11 | Apr 23–26, 2021 | 3,304 |
| 12 | May 28–31, 2021 | 3,280 |
| 13 | June 25–30, 2021 | 3,392 |
| 14 | July 30–Aug 4, 2021 | 3,349 |
| 15 | Aug 27–Sep 1, 2021 | 3,347 |
| 16 | Sep 24–Sep 29, 2021 | 3,310 |
| 17 | Oct 29–Nov 4, 2021 | 3,284 |
| 18 | Nov 26–Dec 1, 2021 | 3,238 |

Note: Observations prior to 11th survey in shaded areas are not included in the sample for this study because questions concerning vaccine were not asked.

**Table 2. Definitions of key variables and its mean, standard errors, and maximum and minimum values.**

| Variables | Definition | Mean | s.d | Max | Min |
|---|---|---|---|---|---|
| *Vaccine_Intention_* | Intention to get a shot of the COVID-19 vaccine. 1 (Very low)–5 (Very high) | 3.88 | 1.33 | 5 | 1 |
| *Subsidy_1000* | It takes 1 if one can receive 1,000 yens subsidy when one gets a shot of the COVID-19 vaccine, otherwise 0. | 0.33 | 0.47 | 1 | 0 |
| *Subsidy_5000* | It takes 1 if one can receive 5,000 yens subsidy when one gets a shot of the COVID-19 vaccine, otherwise 0. | 0.33 | 0.47 | 1 | 0 |
| *No_vaccine* | It takes 1 if one has been vaccinated, otherwise 0. | 0.69 | 0.46 | 1 | 0 |
| *Rate_vaccine* | Rate of vaccinated people in the residential prefecture and the time point of the survey. | 0.24 | 0.27 | 0.66 | 0.01 |
| *Self_motivation* | In deciding whether to get the shot of the COVID-19 vaccine, it is important to save your life. 1 (strongly disagree)–5 (strongly agree). | 4.18 | 0.67 | 5 | 1 |
| *Prosocial_motivation* | In deciding whether to get the shot of the COVID-19 vaccine, it is important that it prevents the spread of COVID-19. 1 (strongly disagree)–5 (strongly agree). | 4.32 | 0.64 | 5 | 1 |
| *Vaccine_harm* | The COVID-19 vaccination has a negative impact on your health. 1 (strongly disagree)–5 (strongly agree). | 2.76 | 0.74 | 5 | 1 |
| *Vaccine_Ineffective* | The COVID-19 vaccination is ineffective. 1 (strongly disagree)–5 (strongly agree). | 2.80 | 0.63 | 5 | 1 |

Note: Values are calculated based on the balanced panel data.

**Table 3. Dependent variable is *Vaccine Intention*.**
   **FE model using balanced panel sample.**

|  | (1) | (2) | (3) |
|---|---|---|---|
| *Subsidy_1000* | −0.383*** | 0.499*** | 0.360*** |
|  | (0.015) | (0.129) | (0.121) |
| *Subsidy_5000* | −0.169*** | 0.472*** | 0.310*** |
|  | (0.013) | (0.139) | (0.229) |
| *No_vaccine* | −0.236*** | −0.413*** | −0.503*** |
|  | (0.020) | (0.019) | (0.021) |
| *Rate_vaccine* | −0.077** | 0.065* | −0.053 |
|  | (0.036) | (0.035) | (0.035) |
| *No_vaccine ×  Subsidy_1000* |  |  | 0.124*** |
|  |  |  | (0.023) |
| *No_vaccine ×  Subsidy_5000* |  |  | 0.145*** |
|  |  |  | (0.022) |
| *Rate vaccine ×  Subsidy_1000* |  | −0.239*** | −0.074** |
|  |  | (0.020) | (0.029) |
| *Rate vaccine ×  Subsidy_5000* |  | −0.189*** | 0.003 |
|  |  | (0.021) | (0.03) |
| *Self_motivation ×  Subsidy_1000* |  | 0.007 | 0.009 |
|  |  | (0.032) | (0.031) |
| *Self_motivation ×  Subsidy_5000* |  | −0.024 | −0.022 |
|  |  | (0.032) | (0.031) |
| *Prosocial_motivation ×Subsidy_1000* |  | −0.158*** | −0.150*** |
|  |  | (0.032) | (0.032) |
| *Prosocial_motivation ×Subsidy_5000* |  | −0.117*** | −0.108*** |
|  |  | (0.033) | (0.032) |
| *Vaccine_harm ×  Subsidy_1000* |  | −0.011 | −0.019 |
|  |  | (0.020) | (0.020) |
| *Vaccine_harm ×  Subsidy_5000* |  | 0.019 | 0.010 |
|  |  | (0.020) | (0.020) |
| *Vaccine_Ineffective ×  Subsidy_1000* |  | −0.050* | −0.053* |
|  |  | (0.028) | (0.028) |
| *Vaccine_Ineffective ×  Subsidy_5000* |  | −0.014 | −0.017 |
|  |  | (0.030) | (0.030) |
| Adj $R^2$ | 0.54 | 0.54 | 0.54 |
| Obs. | 93,288 | 93,288 | 93,288 |

**Note**: Numbers within parentheses are robust standard errors clustered on prefectures where respondents resided. As the set of independent variables, we included the number of deaths and infected persons in residential prefectures at the time of surveys. However, its results are not reported. Test for cross-terms is;

***$p < 0.01$, **$p < 0.05$, *$p < 0.10$

**Table 4. Dependent variable is *Vaccine Intention*.**
       **FE model using sub-sample of unvaccinated people.**

|  | (1) No vaccine | (2) No vaccine | (3) No vaccine in the last survey | (4)) No vaccine in the last survey |
|---|---|---|---|---|
| *Subsidy_1000* | −0.328*** | 0.413*** | −0.152*** | 0.210 |
|  | (0.016) | (0.099) | (0.023) | (0.170) |
| *Subsidy_5000* | −0.115*** | 0.367*** | 0.072*** | 0.406** |
|  | (0.013) | (0.098) | (0.023) | (0.174) |
| *Rate_vaccine* | −0.184*** | −0.138*** | −0.902*** | −0.812*** |
|  | (0.052) | (0.039) | (0.073) | (0.057) |
| *Rate vaccine × Subsidy_1000* |  | −0.139*** |  | −0.095 |
|  |  | (0.050) |  | (0.073) |
| *Rate vaccine × Subsidy_5000* |  | 0.001 |  | −0.173** |
|  |  | (0.050) |  | (0.074) |
| *Self_motivation × Subsidy_1000* |  | −0.021 |  | −0.019 |
|  |  | (0.021) |  | (0.033) |
| *Self_motivation × Subsidy_5000* |  | −0.054** |  | −0.042 |
|  |  | (0.021) |  | (0.034) |
| *Prosocial_motivation ×Subsidy_1000* |  | −0.140*** |  | −0.098*** |
|  |  | (0.022) |  | (0.035) |
| *Prosocial_motivation ×Subsidy_5000* |  | −0.083*** |  | −0.025 |
|  |  | (0.022) |  | (0.036) |
| *Vaccine_harm × Subsidy_1000* |  | 0.035* |  | 0.107*** |
|  |  | (0.019) |  | (0.035) |
| *Vaccine_harm × Subsidy_5000* |  | 0.051*** |  | 0.051 |
|  |  | (0.019) |  | (0.036) |
| *Vaccine_Ineffective × Subsidy_1000* |  | −0.049** |  | −0.091** |
|  |  | (0.023) |  | (0.041) |
| *Vaccine_Ineffective × Subsidy_5000* |  | −0.017 |  | −0.069 |
|  |  | (0.022) |  | (0.043) |
| Adj $R^2$ | 0.55 | 0.55 | 0.55 | 0.55 |
| Obs. | 64,446 | 64,446 | 13,065 | 13,065 |

**Note**: Numbers within parentheses are robust standard errors clustered on prefectures where respondents resided. As the set of independent variables, we included the number of deaths and infected persons in residential prefectures at the time of the surveys; however, its results are not reported.